\documentclass[submission, Proceedings]{SciPost}

\pdfoutput=1

\usepackage{multirow}
\usepackage{textgreek}
\usepackage{array}

\begin{document}

\begin{center}{\Large \textbf{
 Tau-lepton Physics at the FCC-ee circular e$\boldsymbol{^+}$e$\boldsymbol{^-}$ Collider
}}\end{center}

\begin{center}
Mogens Dam\\[1mm]
Niels Bohr Institute, \\[1mm] Copenhagen University\\[1mm]
\texttt{dam@nbi.dk}\\[2mm]
\today
\end{center}

\definecolor{palegray}{gray}{0.95}
\begin{center}
\colorbox{palegray}{
  \begin{tabular}{rr}
  \begin{minipage}{0.05\textwidth}
    \includegraphics[width=9mm]{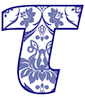}
  \end{minipage}
  &
  \begin{minipage}{0.82\textwidth}
    \begin{center}
    {\it Proceedings for the 15th International Workshop on Tau Lepton Physics,}\\
    {\it Amsterdam, The Netherlands, 24-28 September 2018} \\
    \href{https://scipost.org/SciPostPhysProc.1}{\small \sf scipost.org/SciPostPhysProc.Tau2018}\\
    \end{center}
  \end{minipage}
\end{tabular}
}
\end{center}


\section*{Abstract}
\vspace{-2mm}
{\bf The future FCC-ee collider
  is designed to
  deliver $\boldsymbol{\mathrm{e^+e^-}}$ collisions to study with
  ultimate precision the Z, W, and Higgs bosons, and the top quark. In a
  high-statistics scan around the Z pole, $\boldsymbol{1.3\times 10^{11}}$
  events $\boldsymbol{\mathrm{Z}\to\tau\tau}$ will be produced, the largest
  sample of $\boldsymbol{\tau\tau}$ events foreseen at any lepton collider.
  With their large boost, $\boldsymbol{\tau}$ leptons from Z decays
  are particularly well suited for precision measurements. The focus
  of this report is
  on tests of lepton universality from precision measurement of
  $\boldsymbol{\tau}$ properties and on tests of charged lepton flavour
  violation in Z decays and in $\boldsymbol{\tau}$ decays. In both of these
  areas,
  FCC-ee promises sensitivities well beyond present experimental limits.
}

\vspace{0pt}
\noindent\rule{\textwidth}{1pt}
\tableofcontents\thispagestyle{fancy}

\newpage

\section{Introduction}
\label{sec:intro}

The future 100-km circular collider FCC at CERN is planned to operate, in
its first mode, as an electron-positron machine,
FCC-ee~\cite{TLEPFirstLook,fcc-ee-cdr}.
The FCC-ee is designed to deliver e$^+$e$^-$ collisions to study with the
highest possible statistics the Z, W, and Higgs bosons, and the top quark.
The run plan, spanning 15 years including commissioning, is shown in
Table~\ref{tab:FCC-ee-runplan}.

\begin{table}[b]
  \small
  \begin{center}
    \caption{Run plan for FCC-ee in its baseline configuration with two
      experiments}
    \label{tab:FCC-ee-runplan}
    \vspace{1mm}
    \begin{tabular}{|l|c|c|c|r|} \hline \hline
      Phase & Run duration  & Center-of-mass  &  Integrated   & Event     \\
      & (years)  & Energies (GeV) & Luminosity (ab$^{-1}$) &  Statistics   \\
      \hline
      FCC-ee-Z & 4  & 88--95    & 150   & $3\times 10^{12}$ visible Z decays \\
      FCC-ee-W & 2  & 158--162  &  12   & 10$^8$ WW events                   \\
      FCC-ee-H & 3  & 240       &  5    & 10$^6$ ZH events                   \\
      FCC-ee-tt & 5  & 345--365 &  1.5 & 10$^6$ $\rm t\overline{t}$ events  \\
      \hline \hline
    \end{tabular}
  \end{center}
\end{table}

In its first phase of operation, FCC-ee is
planned to produce $3 \times 10^{12}$ visible Z decays in a scan around the Z
pole, more than five orders of magnitude more than at LEP. This will allow an
extreme precision on the
determination of the Z-boson parameters that are important
inputs to precision tests of the Standard Model (SM). As an important example,
the large data sample, combined with the exquisite determination of the
centre-of-mass energy by resonant depolarization, will allow measurements of
the Z mass and width both to precisions of about 100 keV. Another example,
where the enormous statistics comes in with full power, is the determination of
$\sin^2\theta_\mathrm{W}^\mathrm{eff}$, where, from only the Z-peak measurement
of the forward-backward asymmetry of muon pairs, a factor
$\mathcal{O}(25)$ improvement is expected with respect to the current
precision from all available data.

The enormous Z-boson
statistics, implying $1.3\times 10^{11}$ decays $\text{Z} \to \tau^+\tau^-$,
also opens unique opportunities for precise studies of the $\tau$ lepton.
The $\tau$ lepton is a convenient
probe in the search for Beyond Standard Model (BSM) physics because of the
well-understood mechanisms that govern its production and decay. In the SM, it
is assumed that the electroweak couplings between the three generations of
leptons are universal, and that the
three lepton family numbers are individually conserved. The latter assumption
is violated in the neutral sector by the observation of neutrino
oscillations. Via loop diagrams, this induces also lepton flavour violation
among charged leptons (cLFV). The rates of such processes are however
negligible, so that any observation of cLFV would be an unambiguous signal for
BSM physics (see \emph{e.g.}\ the recent review in Ref.~\cite{Calibbi:2017uvl}).

This report focuses on two main topics: \emph{i})~test of lepton universality
via precision measurements of $\tau$-lepton properties, and \emph{ii})~tests
of cLFV in the decay of Z bosons and in the decay of $\tau$
leptons. Within both areas, FCC-ee promises sensitivities far beyond current
experimental bounds.

To illustrate the potential of FCC-ee for precise $\tau$-physics measurements
it is useful to take a look back at LEP.
Based on rather modest samples of $\mathcal{O}(10^5)$ $\tau\tau$ events,
the LEP experiments were able to take large steps forward in the
measurements of $\tau$-lepton properties, in particular of the lifetime and
branching fractions.
Since then,
the $\mathcal{O}(10^3)$ times larger statistics at the $b$-factories has allowed an
improvement in the lifetime measurement by a relatively modest factor of three,
whereas most LEP branching fraction measurements, in particular those for the
leptonic final states, still stand unchallenged. The advantage of a Z factory
relative to a $b$ factory lies in the nine times higher boost of
the $\tau$s. With the increased flight distance the lifetime measurement
becomes easier, but the higher boost also has strong positive effects on the
quality of the particle identification of the final state particles.

Also for the tests of cLFV in Z decays, LEP measurements still stand largely
unchallenged.



\section{FCC-ee and Detectors}
\label{sec:instruments}

\begin{figure}
  \centering
  \includegraphics[width=0.43\textwidth]{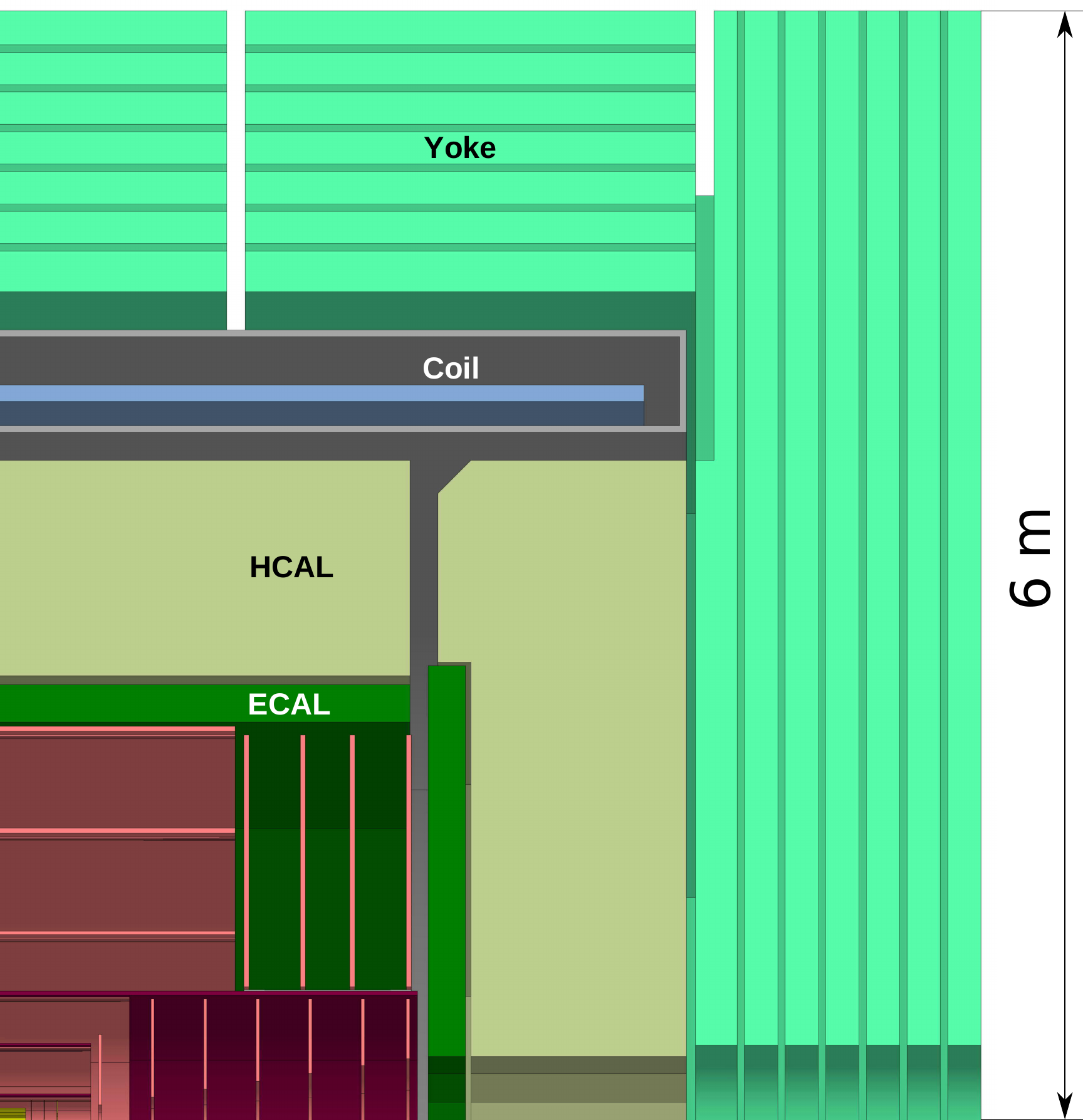}
  \hspace{5mm}
  \includegraphics[width=0.46\textwidth]{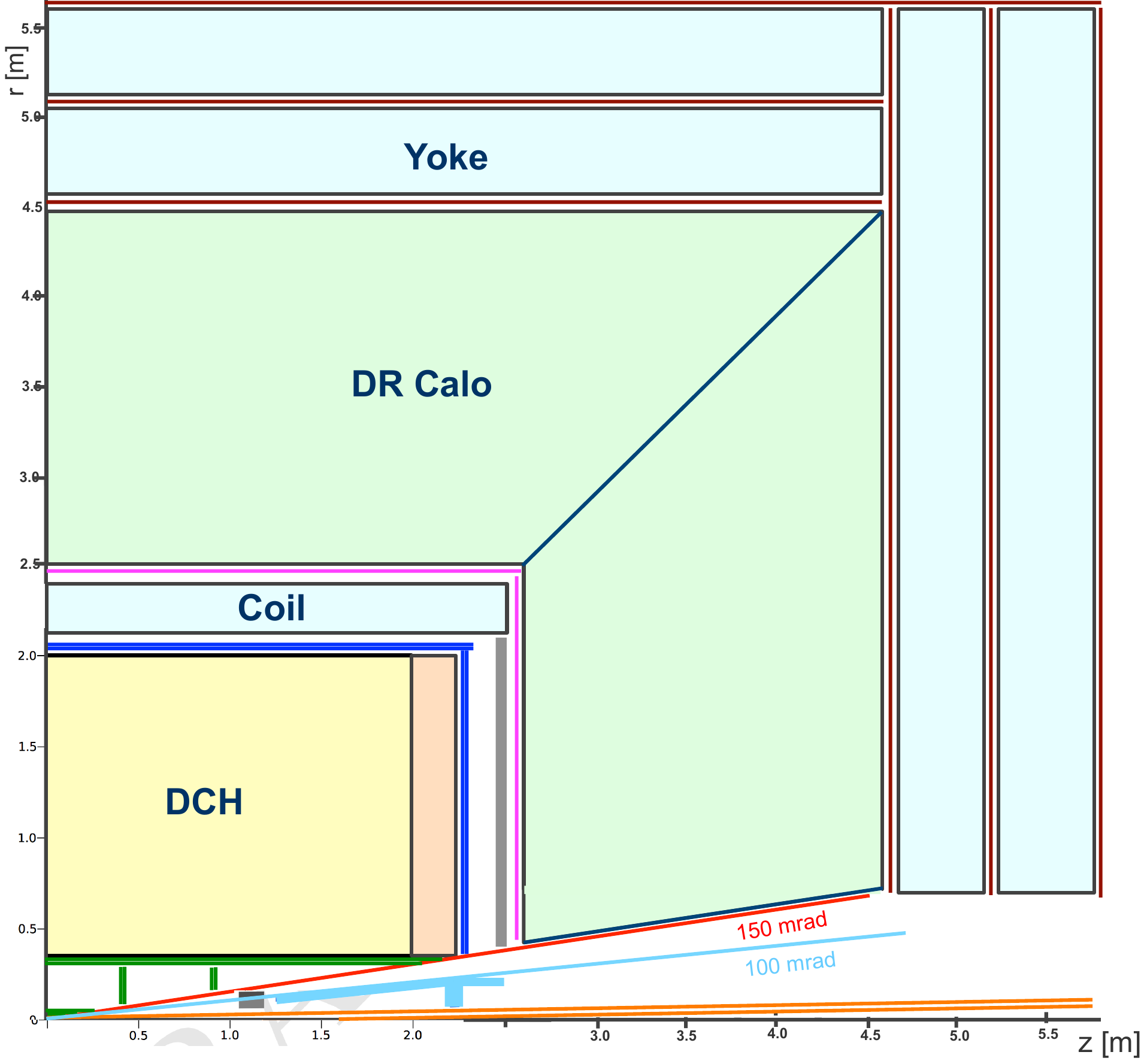}
  \caption{Vertical cross sections showing the top right quadrants of CLD
    (left) and IDEA (right)}
  \label{fig:detectors}
\end{figure}

The unrivalled luminosity performance of FCC-ee is achieved via the use of
techniques inspired from $b$-factories: strong focussing of very low emittance
beams combined with full-energy top-up injection into separate e$^+$ and e$^-$
rings. Circular colliders have the advantage of delivering collisions to
multiple interaction points, which allows different detector designs to be
studies and optimised. In the current design, FCC-ee has two interaction points,
and two complementary detector concepts have been studied~\cite{fcc-ee-cdr}:
\mbox{\textit{i}) CLD}, a consolidated option based on the detector design developed
for CLIC, with a silicon tracker and a 3D-imaging highly-granular calorimeter
system; and \textit{ii}) IDEA, a bolder, possibly more cost-effective, design,
with a short-drift wire chamber and a dual-readout
calorimeter. Cross-sectional views of the two detector concepts are shown in
\mbox{Figure\ \ref{fig:detectors}}.

Focus for the detector-design effort is an
extreme control of systematic effects matching, as far as possible, the
supreme statistical precision of the physics samples, in particular of the
very large Z-boson sample.
Both detector concepts feature a 2 Tesla solenoidal magnetic field (limited in
strength by the 30~mrad beam crossing angle and the requirement of keeping
the beam emittance very low), a small pitch, thin layers vertex detector
providing an excellent impact parameter resolution for lifetime 
measurements,
a highly transparent tracking system providing a superior momentum resolution,
a finely segmented calorimeter system with excellent energy resolution for
$\text{e}/\gamma$ and hadrons,
and a very efficient muon system.
Important figures for the detector performance are summarized in
\mbox{Table \ref{tab:DetPerf}}, where they are compared to those of a typical
LEP detector.
\begin{table}[b]
  \setlength{\extrarowheight}{2pt}
  \begin{center}
    \caption{Important performance figures for a typical detector at FCC-ee
      and at LEP}
    \label{tab:DetPerf}
    \vspace{1.5mm}
    \begin{tabular}{|l|l|c|c|c|} \hline \hline
      \multicolumn{3}{|l|}{Quantity} & LEP & FCC-ee \\ \hline \hline
      \multirow{2}{*}{\shortstack[l]{Impact parameter\\[1pt] resolution}} & \multirow{2}{*}{\(\displaystyle \sigma_d = a \oplus
        \frac{b\cdot\text{GeV}}{p_\mathrm{T} \sin^{2/3}\theta} \)} & $a$ &
      20 \textmu{}m & 3 \textmu{}m \\ \cline{3-5}
      &  & $b$ & 65 \textmu{}m & 15
      \textmu{}m \\
      \hline
      \multirow{2}{*}{\shortstack[l]{Momentum\\[2pt] resolution}} & \multirow{2}{*}{\(\displaystyle
        \frac{\sigma(p_\mathrm{T})}{p_\mathrm{T}} =
        \frac{a \cdot p_\mathrm{T}}{\text{GeV}} \oplus b \)} & $a$ & $6\times 10^{-4}$ &
      $2 \times 10^{-5}$ \\ \cline{3-5}
      & & $b$  & $5\times 10^{-3}$ & $1\times 10^{-3}$  \\
      \hline
      \multirow{2}{*}{\shortstack[l]{ECAL energy \\[1pt] resolution}} & \multirow{2}{*}{\(\displaystyle \frac{\sigma(E)}{E} =
        \frac{a}{\sqrt{E/\text{GeV}}} \oplus b \)} & $a$ & 0.2 & 0.15 \\ \cline{3-5}
        & & $b$
        & 0.01 & 0.01 \\
        \hline
        \multicolumn{3}{|l|}{ECAL transverse granularity} &
        $15 \times 15$ mrad$^2$ & $3 \times 3$ mrad$^2$ \\
      \hline \hline
    \end{tabular}
  \end{center}
\end{table}
In particular for the vertexing and tracking performance,
large improvements are observed compared to LEP. The development of ultra-thin,
fine-pitch silicon sensors plays an important role for this development, as does
the application of a smaller beam pipe with a radius of only 15~mm, about a
quarter of that at LEP, allowing for a first measurement of charged particles
very close to the primary vertex. For the calorimeter system, the much
finer granularity is important for the delicate analysis of the
collimated topologies of multibody hadronic tau decays.

\section{Tau Lepton Properties and Lepton Universality}

High precision measurements of the mass, lifetime, and leptonic branching
fractions of the $\tau$ lepton can be used to test lepton universality.

Firstly, the ratio of the weak-charged-current couplings between muons and
electrons, can be derived from the relation
\begin{equation}
  \left( \frac{g_\mu}{g_\mathrm{e}} \right)^2 =
  \frac{\mathcal{B}(\tau\to\mu\bar{\nu}\nu)}{\mathcal{B}(\tau\to\mathrm{e}\bar{\nu}\nu)} \cdot
  \frac{f_{\tau\mathrm{e}}}{f_{\tau\mu}} ,
\end{equation}
where the phase-space factors are
$f_{\tau\mathrm{e}} = 1$ and $f_{\tau\mu} = 0.97254$.
Current data support universality to the
precision $\delta(g_\mu/g_\mathrm{e}) = 0.14$\%~\cite{Lusiani:tau2018}.

\begin{figure}
  \begin{center}
    \includegraphics[width=0.48\textwidth]{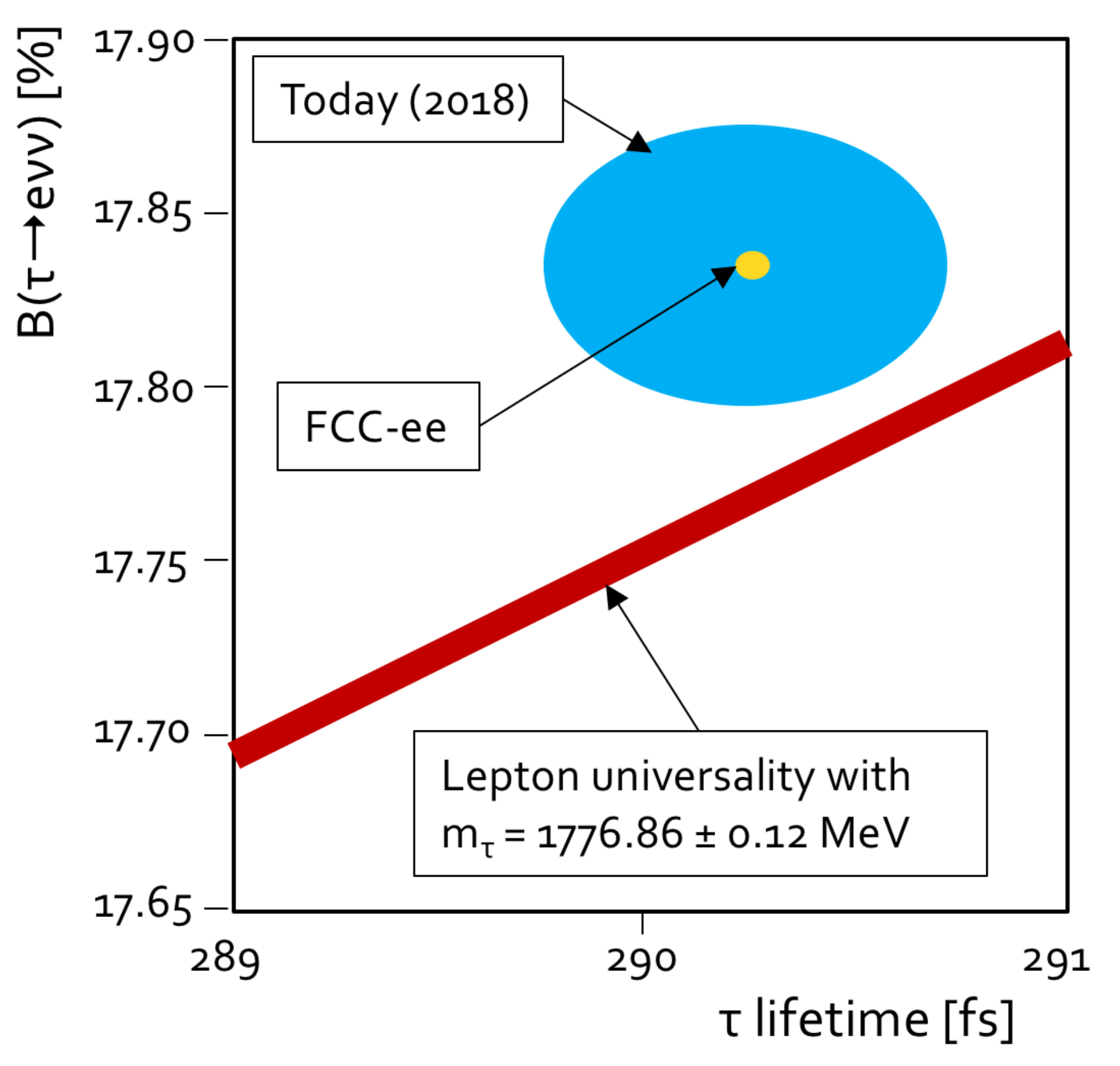}
    \caption{Branching fraction of $\tau \to \rm e \bar{\nu}\nu$
      versus $\tau$ lifetime. The current world averages of the direct
      measurements are indicated with the blue ellipse. Suggested FCC-ee
      precisions are provided with the small yellow ellipse (central
      values have been arbitratily set to todays values). The Standard
      Model functional dependence of the two quantities, depending on the
      $\tau$ mass, is displayed by the red band.}
    \label{fig:tauLFU}
  \end{center}
\end{figure}

Secondly, the ratio of the weak-charged-current couplings between $\tau$ and
electron and between $\tau$ and muon can be derived from the relation
\begin{equation}
  \left( \frac{g_\tau}{g_\ell} \right)^2 =
  \frac{\mathcal{B}(\tau\to\ell\bar{\nu}\nu)}{\mathcal{B}(\mu\to\ell\bar{\nu}\nu)} \cdot
  \frac{\tau_\mu m_\mu^5}{\tau_\tau m_\tau^5} \cdot
  \frac{f_{\mu\mathrm{e}}}{f_{\tau\ell}} \cdot
  \frac{R_\gamma^\mu R_\mathrm{W}^\mu}{R_\gamma^\tau R_\mathrm{W}^\tau} ,
\end{equation}
with $\ell = \mathrm{e}, \mu$, and $f_{\mu\mathrm{e}} = 0.99981$, and where
the last factor represents small radiative and electroweak
corrections~\cite{PhysRevLett.61.1815}.
Current data support universality to the
precision $\delta(g_\tau/g_\mathrm{e}) \simeq \delta(g_\tau/\mu) = 0.15$\%%
~\cite{Lusiani:tau2018}, with the uncertainty dominated by the
measurement of the $\tau$ leptonic branching fractions and lifetime.
Figure~\ref{fig:tauLFU} shows
the current world-average situation for the $\tau\to\mathrm{e}\bar{\nu}\nu$
universality test. Also shown is the situation after a suggested one order of
magnitude improvement in the $\tau$ branching fraction and lifetime
measurements which is within reach at FCC-ee, as discussed below.
At this level of precision, the universality test would be limited by the mass
measurement, if no new measurements would be available. While FCC-ee may
possibly be able to improve the $m_\tau$ measurement by a small factor,
substantial improvements are more likely to come from a next generation of
$\tau$-factory experiments at the production threshold.

\subsection{Lifetime}

The world-average value of the $\tau$-lepton lifetime is
$\tau_\tau = 290.3\pm 0.5$~fs~\cite{PDG2018}.
Precision measurements
were pioneered by the LEP experiments in the early 1990'ies following the
deployment of their precise silicon vertex
detectors~\cite{ALEPH_TauLT97,DELPHI_TauLT04,L3_TauLT00,OPAL_TauLT96}. More
recently, Belle, with its $\mathcal{O}(10^3)$ times larger statistics,
has improved on these measurements~\cite{Belle_TauLT14}.

The
single most precise measurement from LEP, 
$\tau_\tau = 290.0 \pm 1.4\ \text{(stat.)} \pm 1.0\ \text{(syst.)}$~fs,
was provided by DELPHI~\cite{DELPHI_TauLT04}.
The analysis employed several complementary methods. The method with the
smallest systematic uncertainty (1.3~fs) was the so-called decay vertex
method, where the flight-distance was measured for $\tau$ decays to three
charged particles. Here, the largest systematic uncertainty (1.0~fs)
came from the 7.5~\textmu{}m accuracy of the vertex detector alignment. This was
estimated from samples of hadronic Z decays with three tracks in one
hemisphere, and its value resulted from the (limited) statistical power of the
test samples.

The Belle measurement,
$\tau_\tau = 290.17 \pm 0.53\ \text{(stat.)} \pm 0.33\ \text{(syst.)}$~fs,
was based on events in which both $\tau$s decayed
to three charged particles. In these events, the constrained kinematics
combined with the longitudinal boost of the $\tau\tau$ system provided by the
asymmentric KEKB collider allowed Belle to reconstruct the two secondary
vertices as well as the primary vertex and this way to extract the flight
distances. As for DELPHI, the dominant systematic uncertainty
was the accuracy of the vertex detector alignment. The
assigned value of 0.3~fs corresponds to a vertex detector alignment accuracy,
defined as it was done by DELPHI, of 0.25~\textmu{}m, \emph{i.e.}\ a factor 30
better than DELPHI.

The prospects for significantly improved  $\tau$
lifetime measurements at \mbox{FCC-ee} are very good.
Several factors contribute to this:
\emph{i})~Like Belle, the FCC-ee detectors have a \mbox{15-mm} radius beam pipe
allowing the first layer of the vertex detector to go four times closer to
the beam line than at LEP;
\emph{ii})~Based on modern small pitch, thin layer technologies, the
\mbox{FCC-ee} vertex detectors have a space point resolution of only
3~\textmu{}m, four times better than at Belle (and LEP);
\emph{iii})~The much improved statistics compared to LEP (and even
Belle) will allow very precise cross checks and studies of systematic effects
to be carried out.
For the systematic uncertainty, taking, perhaps conservatively, the
0.25~\textmu{}m alignment uncertainty from Belle as an indication of the
achievable precision, this translates immediately, with the higher boost,
into a systematic precison at the level of 0.04~fs.
Relative to LEP, the statistical precison will improve
not only because of the much larger event sample but also due to
the higher sensitivities of the vertex detectors. Hence, a factor
$\mathcal{O}(10^3)$ improvement relative to LEP can be expected resulting in a
statistical uncertainty of 0.001~fs.

Finally, it can be noted that the enormous FCC-ee statistics will
allow extracting the lifetime from numerous complementary
methods with partly different systematics. An obvious option would be, like
Belle, to use events where both $\tau$s decay to three charged particles. At
FCC-ee, nearly $3\times 10^9$ events of this topology will be avaliable.

\subsection{Leptonic Branching Fractions}

Our knowledge about the $\tau$ leptonic branching fractions~\cite{PDG2018},
$\mathcal{B}(\tau\to\mathrm{e}\bar{\nu}\nu) = 17.82 \pm 0.04$\%
and
$\mathcal{B}(\tau\to\mu\bar{\nu}\nu) = 17.33 \pm 0.04$\%,
is comletely dominated by results from $1.3\times 10^5$
$\mathrm{Z}\rightarrow\tau\tau$ events collected at LEP~\cite{ALEPH_TauBf05,DELPHI_TauBfLept99,L3_TauBfLept01,OPAL_TauBfE99,OPAL_TauBfMu03}.
Notwithstanding the $\mathcal{O}(10^3)$ times larger $\tau\tau$ samples at
the $b$-factories, no results on this subject have appeared from there.
Here, therefore, FCC-ee seems unrivalled in the pursuit of
improvements.

Looking back at LEP, ALEPH provided
the single most precise measurement of the leptonic branching fractions with
statistical and systematic uncertainties of 0.070\% and 0.032\%, respectively~\cite{ALEPH_TauBf05}.
With the FCC-ee data sample, the statistical uncertainty will reach a
negligible 0.0001\% level. Even if no improvements
in  detector performance relative to ALEPH would be realized, one would
expect, with the much larger data sample,
also a substantial reduction of the systematic uncertainties. In fact, for the
ALEPH analysis, all major systematics contributions were estimated from
limited-size test samples, and were therefore essentially statistical in
nature. Adding to
this the substantially improved performance of the FCC-ee detectors, in
particular the finer granularity of the calorimeters,
a reduction of the systemtic uncertainty by an order of magnitude
therefore seems within reach, reducing it to the level of 0.003\%.

\subsection{Mass}

The world-average value of the $\tau$ mass is
$m_\tau = 1776.86 \pm 0.12$~MeV~\cite{PDG2018}, with the most precise single measurement
provided by BESIII from an energy scan around the $\tau$-pair production
threshold~\cite{BES3_TauMass14}. Alternatively, $m_\tau$ can be extracted from
the endpoint of the pseudomass distribution from $\tau\to 3\pi\nu$ decays,
where the pseudomass is derived from the four-momentum of
the $3\pi$ system together with the beam energy. This method, originally
pioneered by ARGUS~\cite{ARGUS_TauMass92}, has been pursued by
OPAL~\cite{OPAL_TauMass00} at LEP and by BaBar\cite{BaBar_TauMass09} and
Belle\cite{Belle_TauMass07} at the $b$ factories.
OPAL reached a systematic precision of 1.0~MeV.
With their much larger statistics, Belle and BaBar were able to improve on
this by nearly a factor of three.
The dominant source of systematic uncertainty for all of these measurements
came from the uncertainty on the calibration of the momentum scale. To derive
this, OPAL compared 45.6~GeV tracks from $\text{Z}\rightarrow\text{ee},\,\mu\mu$
events between real and simulated data and extrapolated the results down to
momenta relevant for $\tau\to 3\pi\nu$ decays. With the much larger
samples and the improved momentum resolution at FCC-ee, improvements relative
OPAL can be certainly envisioned. The question is whether
these are large enough to provide a competitive measurement.
The statistical uncertainty will be at a negligible 0.004~MeV level.
For the control of the mass scale one can envisaged to use the decay
$D^+\to K^-2\pi^+$ where the $D^+$ mass is known to
0.050~MeV~\cite{PDG2018}. Hence, a
systematic uncertainty somewhat lower than the current precision,
\emph{e.g.}\ 0.1~MeV, may be within reach. With more than $2\times 10^8$
decays $\tau\to 5\pi\nu$ avalilable, this mode can be likely used for a
complementary measurement.

In summary, FCC-ee may be able to improve somewhat on the \emph{current}
world-average precision of the $\tau$ mass. In all likelihood, however, at the
time of the FCC-ee, larger improvements have been already established
from new experiments at the production threshold.

\section{Charged Lepton Flavour Violation in Z Decays}

Searches for flavour violating Z decays into $\mu$e, $\tau\mu$, and $\tau$e
final states have been performed at LEP and, more recently, at LHC. The
current best bounds are summarized in Table~\ref{tab:CLFVZ}. The LHC results,
which so far all come from ATLAS, are based on about $10^9$ Z decays,
corresponding to around one percent of the total luminosity expected with the
forthcoming high-luminosity upgrade of the LHC.
Hence, future improvements can be expected both from ATLAS and from CMS,
if/when they decide to enter the scene.
\begin{table}[h]
  \setlength{\extrarowheight}{2pt}
  \begin{center}
    \caption{Branching fraction limits (95\% CL) on charged lepton flavour
      violation from Z decays}
    \label{tab:CLFVZ}
    \vspace{1mm}
    \begin{tabular}{|l|lr|lr|} \hline
      Mode & \multicolumn{2}{c|}{LEP} & \multicolumn{2}{c|}{LHC} \\ \hline
      $\text{Z} \rightarrow \mu\text{e}$ & OPAL~\cite{OPAL95_CLFVZ} & $1.7 \times 10^{-6}$ &
      ATLAS~\cite{Aad:2014bca} & $7.5\times 10^{-7}$ \\
      $\text{Z} \rightarrow \tau\mu$ & DELPHI~\cite{DELPHI97_CLFVZ} & $1.2 \times 10^{-5}$ &
      ATLAS~\cite{Aaboud:2018cxn} & $1.3\times 10^{-5}$ \\
      $\text{Z} \rightarrow \tau\text{e}$ & OPAL~\cite{OPAL95_CLFVZ} & $9.8 \times 10^{-6}$ &
      ATLAS~\cite{Aaboud:2018cxn} & $5.8\times 10^{-5}$ \\
      \hline
    \end{tabular}
  \end{center}
\end{table}

\subsection{$\boldsymbol{\text{Z} \rightarrow \mu\mathrm{e}}$}

The decay $\text{Z} \rightarrow \mu\text{e}$ has the particularly clean
signature of a final state with an electron and a muon whose invariant mass
equals the Z mass. At LEP, the search for this mode was \emph{background free},
\emph{i.e.}\ there were no
background events in the ``signal box''. In ATLAS, the search had
several background sources the most important being
$\text{Z} \rightarrow \tau\tau$ events, with the two $\tau$s decaying as
$\tau\to\mathrm{e}\bar{\nu}\nu$ and
$\tau\to\mu\bar{\nu}\nu$, 
and multijet events
where the two final state leptons were either genuine or fake.
Since the ATLAS search has backgrounds, its sensitivity can be expected to
scale as
$1/\sqrt{\mathcal{L}}$, where $\mathcal{L}$ is the
collected luminosity, implying ultimately an order of
magnitude improvement down to sensitivities to branching fractions somewhat
\mbox{below $10^{-7}$}.

With $3\times 10^{12}$ visible Z decays, FCC-ee will be able to improve the
LEP sensitivity considerably for this mode.
At some level, as at LHC, a background from $\text{Z} \rightarrow \tau\tau$
with the two $\tau$s decaying
as $\tau\rightarrow\text{e}\bar{\nu}\nu$ and
$\tau\rightarrow\mu\bar{\nu}\nu$, and where the final state e and $\mu$
are produced at
the end-point of their momentum spectra, will eventually show up.%
\footnote{In principle there is also a background source from
  $\text{Z}\rightarrow\mu\mu$, with one of the muons decaying inside the
  tracking volume. The probability of a 45.6~GeV muon decaying over a distance
  of 2~m is
  $5.6\times 10^{-6}$. However, only a very small fraction of the produced
  electrons come out at the end-point of the momentum spectrum, so this
  background is negligible. It is interesting to notice, that in more than one
  million
  $\text{Z}\rightarrow\mu\mu$ events at FCC-ee one of the muons will decay
  before
  reaching the calorimeter. This potentially opens the possiblity of
  measuring the polarisation of the muons from Z decays in the same way as the
  tau polarisation was measured at LEP in the decays
  $\tau\rightarrow\ell\bar{\nu}\nu$, $\ell=\text{e},\mu$.}
For the
excellent momentum resolution of the FCC-ee detectors, corresponding to
$\sigma_p/p \simeq 1.5 \times 10^{-3}$ for $p=45.6$~GeV, this background
is very small corresponding to a Z-boson branching fraction of about
$10^{-11}$ (with a factor of nearly two included due ot the longitudunal spin
correlation of the two $\tau$s). Here FCC-ee has a large advantage over LHC. Not
only do the FCC-ee detectors have a significant better momentum resolution, but
the Z-mass constraint is also much more powerful. Whereas at LHC, the mass
constraint is limited by the 2.5~GeV natural width of the Z boson, at FCC-ee
the mass is known from the collision energy which has a dispersion of
85~MeV ($0.9\times 10^{-3}$) from the beam energy spread of $1.3 \times
10^{-3}$.

A potentially more serious background arises due to so-called catastrophic
brems\-strahlung of muons in the material of the electromagnetic
calorimeter (ECAL) by which a muon radiates off a significant fraction of its
energy, with the subsequent risk of being
mis-identified as an electron.
NA62 has made detailed studies of this process for their liquid
krypton calorimeter~\cite{NA62_2011}. From this work, one can conclude
that the probability of a 45~GeV muon to deposit more than 95\% of its energy
in the 27~radiation length ($X_0$) deep calorimeter is $4\times 10^{-6}$,
with good agreement (within 10\%) between real and simulated data. If this
number would represent the true FCC-ee propability,
$P_{\mu\mathrm{e}}$, of a muon to be mis-identified as an electron,
a fraction $3\times 10^{-7}$ of all Z decays would go through
$\text{Z}\rightarrow\mu\mu$ and appear mistakenly as e$\mu$ final states,
eventually limiting the sensitivity of the $\text{Z} \rightarrow \mu\text{e}$
search to about an
order of magnitude lower than that, if one assumes that $P_{\mu\mathrm{e}}$
can be controlled to about 10\% of its value. In practice, however,
$P_{\mu\mathrm{e}}$ will likely be somewhat lower than the NA62
number. Firstly, the energy resolution of the FCC-ee ECAL
is about 2.5\% at 45~GeV, and a more stringent 
requirement on the energy deposit than 95\% can be placed. Secondly,
longitudinal 
segmentation of the ECAL allows setting requirements on the energy
deposit in the first few radiation lengths where the
bremsstrahlung propability will be lower than for the full depth. Very
important for this measurement is the ability to precisely determine
$P_{\mu\mathrm{e}}$ from the data themselves. Longitudinal
segmentation of the calorimeter certainly helps to this end, but a truely
independent method of separating electrons and muons would make a more
powerful tool. Even if the e/$\mu$ separation of a d$E/$d$x$ measurement
at $p=45.6$~GeV would not be ideal, it could be still employed to manipulate the
electron-to-muon ratio in test samples in order to study the
calorimeter response. In this respect, it is encouraging to note that the
drift chamber of the IDEA detector, through the use of cluser counting,
promises a e/$\mu$ separation at $p=45.6$~GeV of 3--4 standard
deviations~\cite{FrancoG2018}.

In conclusion, a sensitivity for the $\text{Z} \rightarrow \mu\text{e}$ mode
at the $10^{-8}$ level should be within reach at FCC-ee. An independent method
for e/$\mu$ separation, as that provided by a powerful d$E/$d$x$ measurement,
could potentially improve this sensitivity by one to two orders of magnitude
potentially all the way down to the $10^{-10}$ level.

\subsection{$\boldsymbol{\text{Z} \rightarrow \tau\mu}$ and
  $\boldsymbol{\text{Z} \rightarrow \tau\mathrm{e}}$ }

The searches for $\text{Z} \rightarrow \tau\mu$ and
$\text{Z} \rightarrow \tau\text{e}$ have many similarities and they will be
here treated under one.
All previous searches are characterised by having background events occuring
in the ``signal box'', so that sensitivities scale as
$1/\sqrt{\mathcal{L}}$, where $\mathcal{L}$ is the
collected luminosity. Ultimately, one can thus expect the LHC sensitivities to
approach the $10^{-6}$ level.

In e$^+$e$^-$ collisions, the pursuit for decays
$\text{Z} \rightarrow \tau\mu\ (\tau\text{e})$ amounts to a search for
events with
a \emph{clear tau decay} in one hemisphere recoiling against
a \emph{beam-momentum muon (electron)} in the other. To illuminate the
analysis, we will investigate these two terms.

Firstly, let us
consider the term \emph{clear tau decay}. Here, the point is to restrict the
analysis to those $\tau$-decay modes, where the probability is minimal of
misidentifying a final-state lepton from $\text{Z}\to\mu\mu$
or $\text{Z}\to\text{ee}$ as a $\tau$ decay.
This immediately excludes the
leptonic modes $\tau\to\mu\bar{\nu}\nu$ and $\tau\to\text{e}\bar{\nu}\nu$. For
the remaining hadronic modes, there may be a
risk of mis-identifying either a muon or an electron as a pion, so that the
decay $\tau\to\pi\nu$ may also have to be excluded, at least for large
$\pi$ momenta. To reach very high purities, it may be ultimately necessary to
restrict the analysis to reconstructed exclusive modes such as
$\tau\to\rho\nu\to\pi\pi^0\nu$ and decays to three (or more) charged particles.

\begin{figure}
  \begin{center}
    \includegraphics[width=0.45\textwidth]{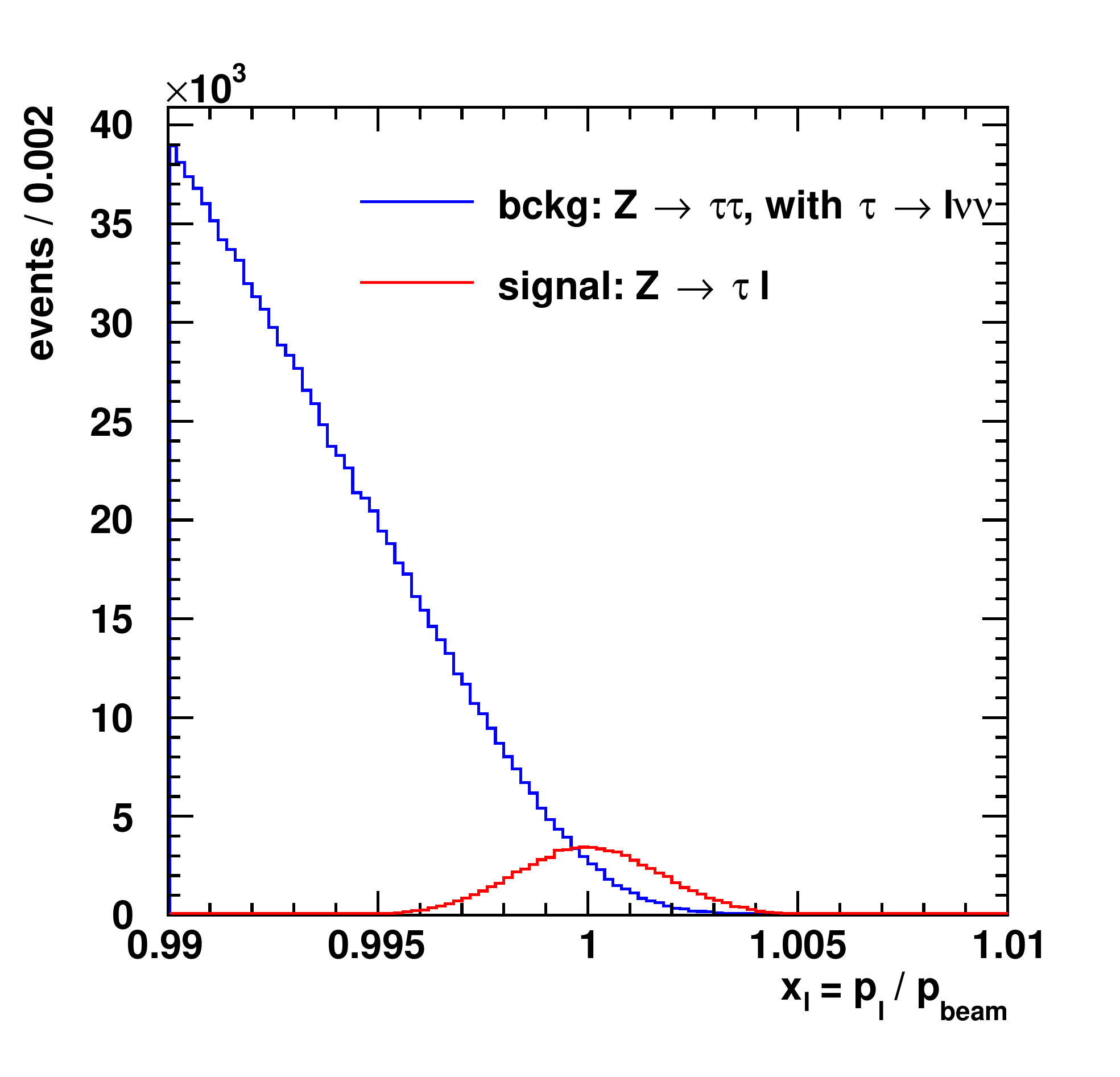}
    \caption{FCC-ee search for the lepton flavour violating decay
      $\text{Z}\rightarrow\tau\ell$, $\ell=\mathrm{e},\mu$.
      Momentum distribution of the final state lepton $\ell$ for the signal
      (red) and for the background from
      \mbox{$\text{Z}\rightarrow\tau\tau$}, with $\tau\rightarrow\ell\bar{\nu}\nu$
      (blue).
      The shown momentum resolution of
      $1.8 \times 10^{-3}$ results from the combination of the
      spread on the collision energy ($0.9\times 10^3$) and the detector
      resolution ($1.5\times 10^{-3}$).
      For illustration, the LVF branching fraction is set here to
      \mbox{$\mathcal{B}(\text{Z}\to\tau\ell) = 10^{-7}$.}}
    \label{fig:ZLFV}
  \end{center}
\end{figure}

Secondly, we have to
consider the term \emph{beam-momentum muon (electron)}. Since neutrinos
are (nearly) massless, the muon (electron) momentum distribution in
$\tau\rightarrow\mu\bar{\nu}\nu$ ($\tau\rightarrow\text{e}\bar{\nu}\nu$)
decays will have
an end-point at the beam momentum. Ignoring the mass of the final state
charged lepton, the muon (electron)
momentum distribution is given by the
expression~\cite{tsaiTau1971},
\begin{equation}
  \frac{1}{\Gamma} \frac{\mathrm{d}\Gamma}{\mathrm{d}x} =
  \frac{1}{3} \left[\left(5-9x^2+4x^3\right) +
    P_\tau \left(1-9x^2+8x^3\right) \right],
\end{equation}
where $x=p/p_\mathrm{beam}$, and $P_\tau$ is the longitudinal polarisation of the
$\tau$ leptons. Hence, the density of events close to the endpoint
depends on $P_\tau$ (and thus on $\sin^2\theta_\mathrm{W}^\mathrm{eff}$),
which is here set to the value $P_\tau=-0.15$, consistent with the LEP
result~\cite{EWWG2005}. The separation of signal and background now depends
on the experimental precision by which a \emph{beam-momentum particle} can be
defined. This is illustrated in Figure~\ref{fig:ZLFV}, where the indicated
momentum spread of $1.8\times 10^{-3}$ arises as a combination of the
$0.9\times 10^{-3}$ spread of the collision energy and the $1.5\times 10^{-3}$
momentum resolution typical for a FCC-ee detector at $p=45.6$~GeV.
By defining the signal box by the
simple requirement $x>1$, it was found that FCC-ee, with the quoted
resolution and a signal efficiency of 25\%, has a
sensitivity to branching fractions down to $10^{-9}$. The
sensitivity scales linearly in the momentum resolution.

\section{Charged Lepton Flavour Violation in $\boldsymbol{\tau}$ Decays}

\begin{figure}
  \centering
  \hspace{1mm}
  \includegraphics[width=0.7\textwidth]{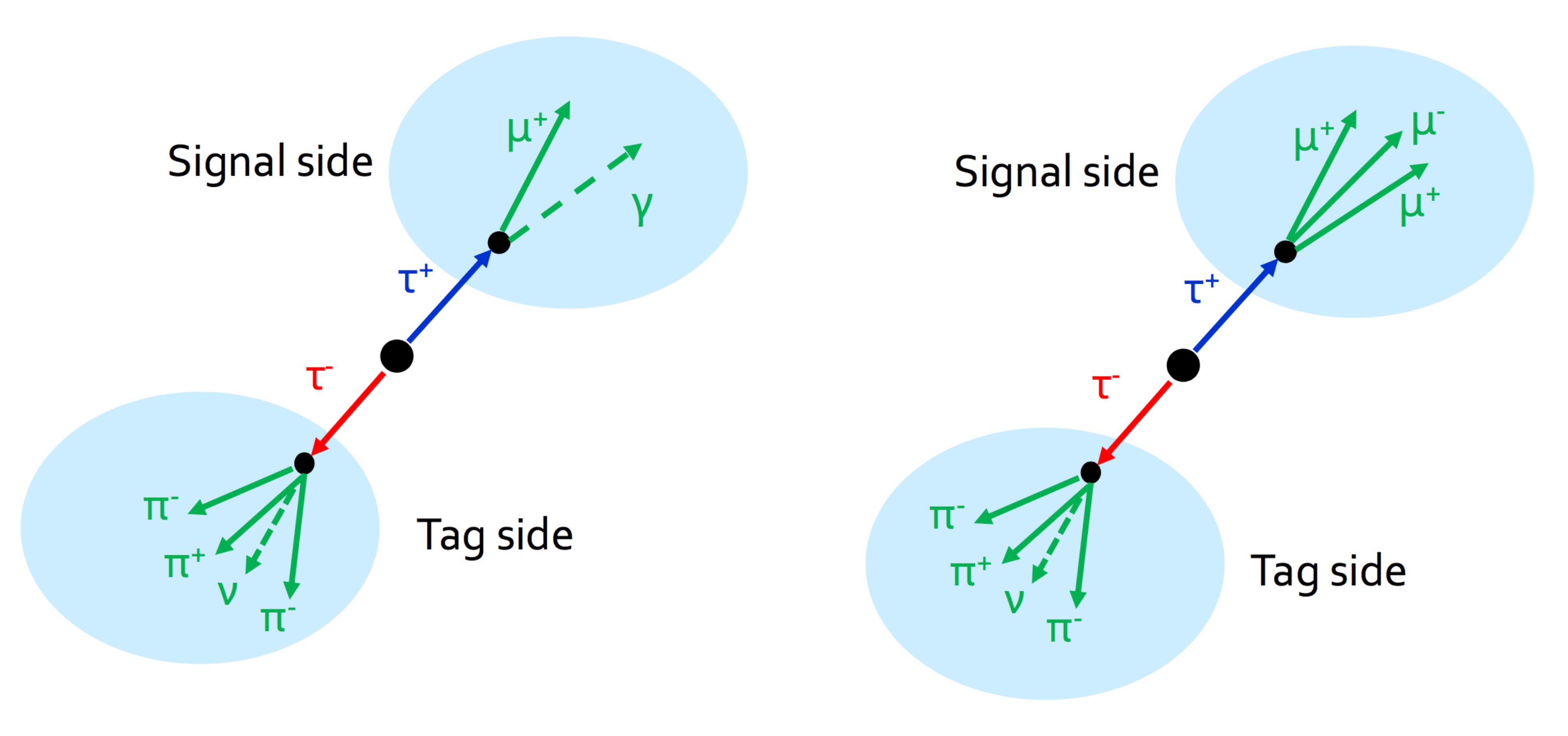}
  \caption{Illustration of the search for lepton flavour violating $\tau$
    decays}
  \label{fig:tau_cLFV}
\end{figure}

Very stringent tests of cLFV have been
performed in muon decay experiments where branching fraction limits below
$10^{-12}$ on both of the decay modes
$\mu^-\to\mathrm{e}^-\gamma$~\cite{MEG_MuToEGamma16}
and $\mu^+\to\mathrm{e^+e^+e^-}$~\cite{SINDRUM_MuTo3E88}
have been established. All models
predicting cLFV in the muon sector imply a violation also in the $\tau$ sector,
whose strength is often enhanced by several orders of magnitude, usually by
some power in the tau-to-muon mass ratio. Studying cLFV processes in $\tau$
decays offers several advantages compared to muon decays. Since the $\tau$
is heavy, more cLFV processes are kinematically allowed. In addition to the
modes $\tau\to\mu/\mathrm{e}+\gamma$ and
$\tau\to\mu/\mathrm{e}+\ell^+\ell^-$, cLFV can be also studied in several
semileptonic modes. The expected $2.6\times 10^{11}$ $\tau$s produced at
FCC-ee exceed the projected Belle~II (50~ab$^{-1}$) statistics by a factor of about
three, raising the possibility that FCC-ee may provide competitive
sensitivities.
The focus here is on $\tau\to 3\mu$ and
$\tau\to\mu\gamma$ as benchmark modes for
evaluating the sensitivity to cLFV. The analysis strategy is illustrated in
Figure~\ref{fig:tau_cLFV}, with a \emph{tag side} to identify a
clear standard-model tau decay and a \emph{signal side} where cLFV
decays are searched for. The present $\mathcal{O}(10^{-8})$ bounds on both
modes are set at the $b$ factories
~\cite{BELLE_TauTo3Lept10,BABAR_TauToLeptGam10}.
As detailed below, about two (one) orders of magnitude improvement can be
expected at FCC-ee for the decay $\tau\to 3\mu$
($\tau\to\mu\gamma$).
This turns out to be
largely compatible with the recently published estimates for
Belle~II~\cite{BELLEII_PhysicsBook}.

\subsection{$\boldsymbol{\tau\to 3\mu}$}

The present bound of $2.1 \times 10^{-8}$ on the $\tau\to 3\mu$ mode comes
from Belle~\cite{BELLE_TauTo3Lept10}.
With the excellent FCC-ee invariant mass
resolution, the search for this mode is expected to be essentially
background free, and a sensitivity down to a branching fractions
of $\mathcal{O}(10^{-10})$ should be within reach.

\subsection{$\boldsymbol{\tau\to\mu\gamma}$}

\begin{figure}
  \begin{center}
    \includegraphics[width=0.95\textwidth]{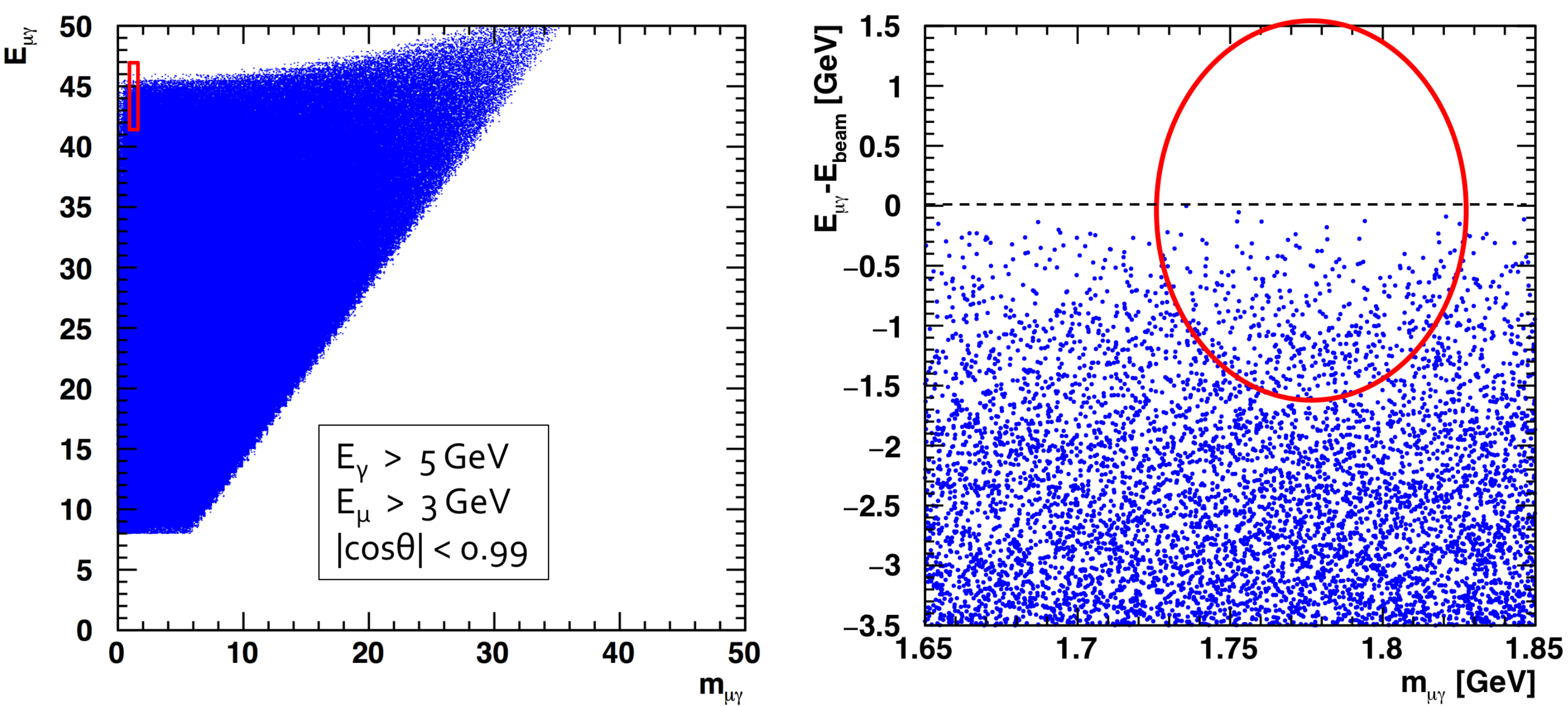}
    \caption{Main background to the $\tau\to\mu\gamma$ search at
      FCC-ee. Reconstructed energy versus mass of all $\mu\gamma$ combinations
      in simulated background events
      $\mathrm{e^+e^-}\to\text{Z}\to\tau^+\tau^-\gamma$
      with one $\tau\to\mu\bar{\nu}\nu$ decay. The right-hand diagram,
      showing an enlargement of the red boxed region from the left-hand
      diagram, has the signal region shown by a red ellipse.}
    \label{fig:mugamma}
  \end{center}
\end{figure}

The present bound of $2.7 \times 10^{-8}$ on the $\tau\to\mu\gamma$ mode
comes from BaBar~\cite{BABAR_TauToLeptGam10}.
The search for this mode is limited by backgrounds, namely radiative events
$\mathrm{e^+e^-}\to\tau^+\tau^-\gamma$,
with one $\tau\to\mu\bar{\nu}\nu$ decay, and the invariant mass of
a $\mu\gamma$ pair accidentially in the signal region. An experimental
study of the signal and this dominant background has
been performed for FCC-ee conditions using the Pythia8 event
generator~\cite{pythia64,pythia81}, and
smearing the output four-vectors with realistic detector resolutions.
The decay mode $\tau\to\pi\nu$ was used to fake the signal with the final state
particles renamed as $\pi\to\mu$ and $\nu\to\gamma$. For the
detector-performance numbers from Table~\ref{tab:DetPerf}, and by further
assuming a position resolution for photons of
$\sigma_x = \sigma_y = (6\ \text{GeV}/E\, \oplus\, 2)$~mm, resolutions on the
mass\footnote{%
In order to de-correlate the $E_{\mu\gamma}$ and $m_{\mu\gamma}$
variables, the mass variable used here is defined by
$m_{\mu\gamma} = m_{\mu\gamma}^\mathrm{raw} \cdot
(E_\mathrm{beam}/E_{\mu\gamma})$,
where $m_{\mu\gamma}^\mathrm{raw}$ is the measured ``raw'' mass.%
}
and energy of $\mu\gamma$ pairs of $\sigma(m_{\mu\gamma}) =
26$~MeV and $\sigma(E_{\mu\gamma}) = 850$~MeV were
derived. Figure~\ref{fig:mugamma} shows the
background events in the $E_{\mu\gamma}$ vs.\ $m_{\mu\gamma}$ plane, with
a signal region corresponding to $2\sigma$ resolutions indicated in the
right-hand plot by the red
ellipse. The statistics in the figure corresponds to nearly $7\times 10^{10}$
visible Z decays or about 2\% of the full FCC-ee statistics. Scaling to the
full statistics, about 20,000 background combinations are expected inside the
signal region from which a sensitivity down to branching fractions of
$2\times10^{-9}$ can be estimated.

As seen from Figure~\ref{fig:mugamma}, the density of background events rises
away from the (dashed) \mbox{$E_{\mu\gamma}-E_\mathrm{beam}=0$} line. In fact,
the rise is linear, so that the 
sensitivity
scales linearly in $\sigma(E_{\mu\gamma})$ and therefore also in the ECAL
energy resolution which is the dominant contribution to $\sigma(E_{\mu\gamma})$.

\section{Summary and Conclusions}

Among all proposed future lepton colliders,
FCC-ee, with
$3\times 10^{12}$ visible Z decays, offers the largest sample of $\tau\tau$
events. From the LEP experience, we know that a Z factory is particularly well
suited for precision $\tau$-physics measurements. Indeed, even today, after
$\mathcal{O}(10^3)$ times more $\tau$ decays have been collected at the
$b$ factories, several LEP measurements still stand unchallenged.
With more than five orders of magnitude more events at FCC-ee than at LEP,
large steps forward in terms of precision can be therefore foreseen. Whereas
almost all LEP measurements were statistics limited, at FCC-ee,
systematic effects will be dominant.
At this stage, systematic uncertainties are hard to estimate,
and all values given here should be taken with this in mind.

A summary of projected measurements of $\tau$-lepton properties is
presented in Table~\ref{tab:SummaryTauProperties}. More than one order of
magnitude improvements are expected in the lifetime and branching fraction
measurements. This will enable tests of lepton universality
down to a precision at the 0.01\% level. Going beyond that, will require also an
improved precision on the $\tau$ mass. Here, FCC-ee may be able to touch upon
the present precision, however, substantial
improvements will be more likely obtained via a high statistics scan of the
production threshold at a next generation $\tau$ factory.
\begin{table}[t]
  \begin{center}
    \caption{Measurement of $\tau$-lepton properties at FCC-ee, compared to
    the present precisions}
    \label{tab:SummaryTauProperties}
    \vspace{1mm}
    \begin{tabular}{lccc} \hline 
      Observable & Present & FCC-ee & FCC-ee 
      \\
      & value $\pm$ error & stat. & syst. 
      \\ \hline 
      $m_\tau$ (MeV) & $1776.86 \pm 0.12$ & 0.004 & 0.1 
      \\ 
      $\mathcal{B}(\tau\to\mathrm{e}\bar{\nu}\nu)$ (\%) & 17.82 $\pm$ 0.05 &
      0.0001 & 0.003 
      \\ 
      $\mathcal{B}(\tau\to\mu\bar{\nu}\nu)$ (\%) & 17.39 $\pm$ 0.05 &
      0.0001 & 0.003 
      \\ 
      $\tau_\tau$ (fs) & 290.3 $\pm$ 0.5 & 0.001 & 0.04 
      \\ \hline 
    \end{tabular}
  \end{center}
\end{table}

\begin{table}[b]
  \begin{center}
    \caption{FCC-ee sensitivities to cLFV
      processes in Z decays and in two benchmark $\tau$-decay modes,
      compared to present bounds. The range of values for the
      $\text{Z}\to\mu$e mode reflects whether or not particle identification
      via d$E/$d$x$ will be available.
    }
    \label{tab:SummaryCLFV}
    \vspace{2mm}
    \begin{tabular}{lcc} \hline
      Decay & Present bound & FCC-ee sensitivity
      \\ \hline
      $\text{Z}\to\mu\mathrm{e}$  & $0.75\times 10^{-6}$ & $10^{-10}$--$10^{-8}$ \\
      $\text{Z}\to\tau\mu$        & $12\times 10^{-6}$   & $10^{-9}$ \\
      $\text{Z}\to\tau\mathrm{e}$ & $9.8\times 10^{-6}$  & $10^{-9}$ \\ \hline
      $\tau\to\mu\gamma$          & $4.4\times 10^{-8}$  & $2\times 10^{-9}$ \\
      $\tau\to3\mu$               & $2.1\times 10^{-8}$    & $10^{-10}$ \\
      \hline
    \end{tabular}
  \end{center}
\end{table}

Precise tests of charged lepton flavour violation in Z decays, with branching
fraction limits in the range $10^{-6}$--$10^{-5}$, were performed
at LEP. LHC measurements have been able to improve on this, so far, by about a
factor of two for the $\text{Z}\to\mu$e channel. In general about an order of
magnitude improvements can be ultimately expected from the full LHC samples.
%
With its enormous Z statistics,
FCC-ee will be able to improve dramatically on this with
sensitivities about four orders of magnitude higher than today's.
The situation is summerized in Table~\ref{tab:SummaryCLFV}.

Very precise tests
of charged lepton flavour violation in
$\tau$ decays have been carried out at the $b$ factories, and will
be further improved at Belle~II. With the larger sample of $\tau$
decays, and with the higher boost, FCC-ee will be able perform competitive
measurements in this area. The situation is
presented in Table~\ref{tab:SummaryCLFV} for two benchmark processes.

\section*{Acknowledgements}

It is my pleasure to thank Olya Igonkina and the Organizing Committee for
giving me the chance to present these studies in the Workshop. Also I would
like to acknowledge them for the excellent organization of the Workshop in the
beautiful venue of the Vondelkerk, Amsterdam.





{\raggedright
\bibliography{tau2018_FCCee.bib}
}

\nolinenumbers

\end{document}